\documentclass{PoS}

\title{DARWIN: \\dark matter WIMP search with noble liquids}

\ShortTitle{DARWIN: dark matter WIMP search with noble liquids}

\author{\speaker{Laura Baudis}\thanks{DARWIN Project Coordinator; on behalf of the DARWIN consortium.}\\
        Physik Institut, University of Zurich\\
        E-mail: \email{laura.baudis@physik.uzh.ch}}

\abstract{DARWIN (DARk matter WImp search with Noble liquids) is an R\&D and design study towards the realization of a multi-ton scale dark matter search facility in Europe, based on the liquid argon and liquid xenon time projection chamber techniques. Approved by ASPERA in late 2009, DARWIN brings together several European and US groups working on the existing ArDM, XENON and WARP experiments with the goal of providing a technical design report for the facility by early 2013. DARWIN will be designed to probe  the spin-independent WIMP-nucleon cross section region below 10$^{-47}$cm$^2$ and to provide a high-statistics measurement of WIMP interactions in case of a positive detection in the intervening years. After a brief introduction,  the DARWIN goals, components, as well as its expected physics reach will be presented.}

\FullConference{Identification of Dark Matter 2010-IDM2010\\
		July 26-30, 2010\\
		Montpellier France}

\newenvironment{packed_enum}{
\begin{enumerate}
  \setlength{\itemsep}{1pt}
  \setlength{\parskip}{0pt}
  \setlength{\parsep}{0pt}
}{\end{enumerate}}

\begin{document}

\section{Introduction}
\label{introduction}

One of the most exciting topics in physics today is the nature of Dark Matter in the Universe. Although indirect evidence for cold dark matter is well established, its true nature is not yet known. The most promising explanation is Weakly Interacting Massive Particles (WIMPs), for they would naturally lead to the observed abundance and they arise in many of the potential extensions of the Standard Model of particle physics. WIMPs could be detected directly by their collisions with nuclei in underground experiments, such a discovery would be a milestone in physics. However, since the predicted signal rates are much lower than one interaction per kg of target material and day, large detector masses and ultra-low backgrounds are necessary ingredients of any experiment aiming to discover WIMPs. 
Results from noble liquid detectors have recently shown that these detectors are among the most promising technology to push the sensitivity of direct WIMP searches far beyond existing limits into the regime of favored theoretical predictions. Liquid argon (LAr) and xenon (LXe), having high charge and light yields for nuclear recoils expected from WIMP-nucleus  scattering, are excellent WIMP targets. A noble liquid Time Projection Chamber (TPC) can offer a scalable, large, self-shielding, homogeneous and position sensitive WIMP detector. The relative size of the charge and light signals, as well as their timing allows efficient discrimination against electron recoil events, and good spatial resolution allows the identification of the neutron background. 

The XENON10 \cite{xenon10} and WARP (2.3 liter) \cite{warp2.3} experiments at the Gran Sasso Underground Laboratory (LNGS) and the ZEPLIN-III experiment \cite{zeplin3} at the Boulby laboratory were successful demonstrators, reaching competitive limits on both spin-independent and spin-dependent WIMP-nucleon cross sections. The XENON100 experiment is taking science data at LNGS, and has published first results from a commissioning run in late 2009 \cite{xenon_PRL105}. The WARP-140 experiment, a 100kg-scale  LAr TPCs, is under commissioning at LNGS, while the ArDM-1t \cite{ardm} detector, with a LAr WIMP target of 850 kg, is under commissioning at CERN, with the goal of underground installation in the Canfranc Underground Laboratory in 2011. While the aimed sensitivities are around 10$^{-45}$cm$^2$ for the spin-independent WIMP-nucleon cross section, a few events per year would be detected for a  cross section at the level of 10$^{-44}$cm$^2$. The proposed XENON1T detector \cite{xenon1t_tdr}, with a total of 1 ton of LXe in the fiducial volume (2.4\,t total liquid xenon mass), would reach another order of magnitude in sensitivity improvement by 2015.

\section{DARWIN Goals and Consortium}
\label{goals}

\begin{figure}[!h]
\begin{center}
\includegraphics[scale=0.60]{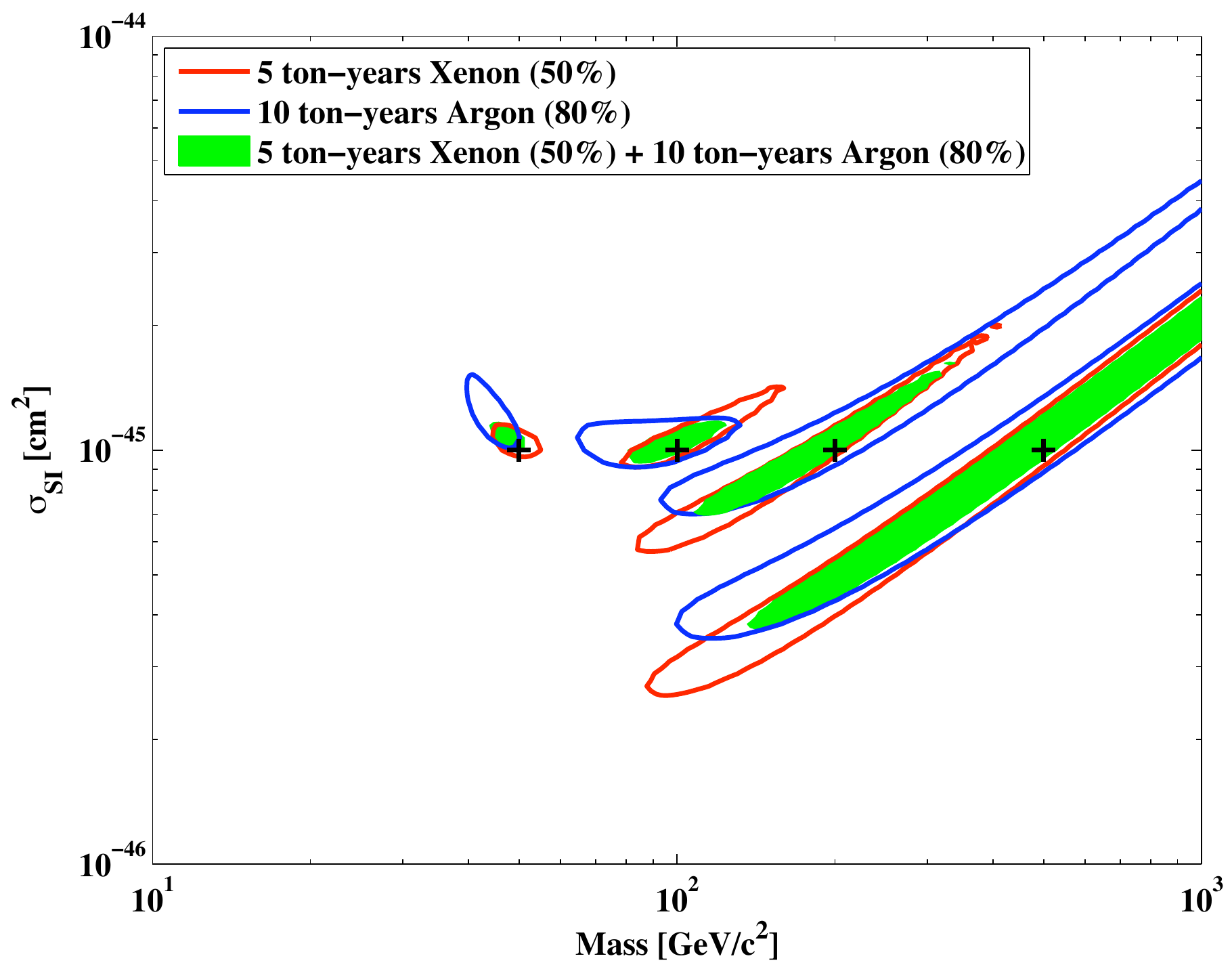}
\caption{\small{Spin-independent WIMP-nucleon cross section versus WIMP mass for a benchmark scenario of 10$^{-45}$cm$^2$ (10$^{-9}$pb) and several WIMP masses. Shown are the 1-$\sigma$ constraints on the WIMP mass for an exposure of 5\,t\,year in LXe (red, with a 50\% acceptance of nuclear recoils) and a 10\,t\,year exposure in LAr (blue, with a 80\% acceptance of nuclear recoils), as well as the improved constraint from the combination of both results (filled regions). }}
\label{fig:sigma_mass}
\end{center}
\end{figure}

DARWIN \cite{darwin} is an R\&D and design study for a multi-ton scale LAr and LXe dark matter search facility, with the goal of probing the cross section region below 10$^{-47}$cm$^2$, or to provide a high-statistics measurement of WIMP interactions in case of a positive detection by one of the aforementioned experiments.  To convincingly demonstrate the dark matter nature of a signal, a measurement of the interaction rate with multiple targets is mandatory. Operating a LAr and a LXe target under similar experimental conditions would allow to measure the dependence of the rate with the target material and hence to better constrain the WIMPs mass (for masses below about 500 GeV/c$^2$, as shown in Figure \ref{fig:sigma_mass}), and to distinguish between spin-independent and spin-dependent couplings ($^{40}$Ar has no spin, while natural xenon contains $^{129}$Xe and $^{131}$Xe, with nuclear spins of the ground states of  1/2$^{+}$  and 3/2$^{+}$ and abundances  of 26.4\% and 21.2\%, respectively). From a technical point of view, there are many common aspects to LAr and LXe dark matter TPCs, starting from the cryostat design, to charge and light readout, to the purification of noble liquids, to the HV systems required for charge drift, uniform field and charge extraction, to the  use of ultra-low radioactivity materials and shields as well as the underground infrastructure and safety aspects.

\begin{figure}[!h]
\begin{center}
\includegraphics[scale=0.50]{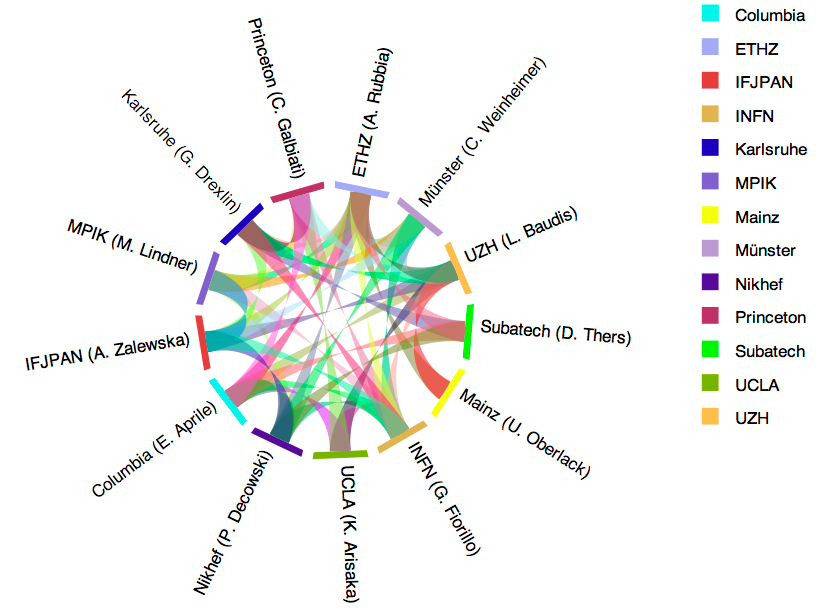}
\caption{\small{The participating institutions in the DARWIN study and the connections among these.}}
\label{fig:consortium}
\end{center}
\end{figure}

Approved by ASPERA \cite{aspera} in late 2009, DARWIN coordinates the European groups active in the noble liquid dark matter search field, bringing in new and valuable expertise from closely related fields such as neutrino physics, as well as associate US partners with long-standing experience in the noble liquid technology.  It thus  unites for the first time the ample expertise in Europe on liquid noble gas detectors, on low-background techniques, on cryogenic infrastructures, on underground infrastructures and shields as well as on the physics related to the direct detection of WIMPs in a coherent manner. Figure \ref{fig:consortium} show the participating institutions and the connection among these.  The connections are established through several work packages (see Section \ref{components} for details), for each group is in general contributing to more than one work package and its sub-tasks.

\section{DARWIN Components}
\label{components}

The DARWIN study has several components, organized into eight different work packages. Here we can list  the main activities only: 

\begin{packed_enum}
\item
Optimization of cryostat and inner TPC design, as well as of the high-voltage and cryogenic systems using performance extracted from real data from several existing LAr and LXe experiments (see Section \ref{introduction}), which we consider as prototypes for the larger scale facility. 
\item
Study of novel, high quantum-efficiency and low-radioactivity light read-out sensors and UV light collection schemes, as well as the light and charge yields of electronic and nuclear recoils at low-energies.
\item
Study of new concepts to read-out the ionization signal, such as large-area thick gas electron multipliers (GEMs), large-area gaseous photomultipliers (GPM) and CMOS pixel detectors coupled to electron multipliers. 
\item 
Study of low-noise, low-power and cost-effective electronics for light and charge read-out, as well as new DAQ and data processing schemes. 
\item
Optimization of noble gas purification procedures concerning traces of water and electronegative impurities, as well as radioactive isotopes such as $^{39}$Ar, $^{85}$Kr and $^{222}$Rn;   studies of material outgassing, liquid handling and purity monitoring procedures. 
\item
Identification of material selection and process control needed for ultra-low background operation; exploration of optimal underground locations and a cost-effective shielding configuration.
\item 
Study of the science impact of this technology and establishment of a framework in which dark matter results from indirect searches, cosmology and the LHC can be combined with direct dark matter searches; asses the impact of potential results on astrophysics.
\end{packed_enum}

The final goal is to deliver a technical design report on the largest scale facility feasible around 2013 (the DARWIN facility) as input for a coordinated proposal for construction and operation of such a detector underground. The scientific performance of such a facility will be determined using a full simulation, including a study of the added value from the combination of data from different WIMP targets, as well as from indirect detection experiments and the LHC.

\section{DARWIN Physics Reach and Timeline}

\begin{figure}[!h]
\includegraphics[scale=0.43]{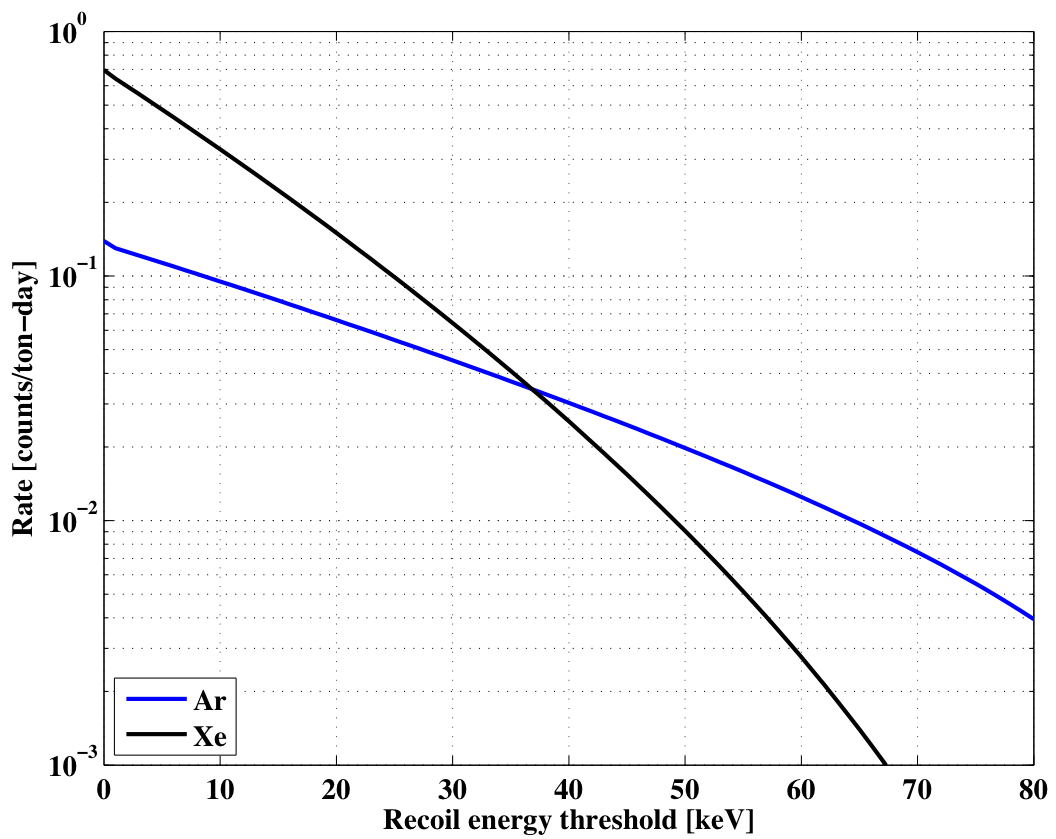}
\includegraphics[scale=0.43]{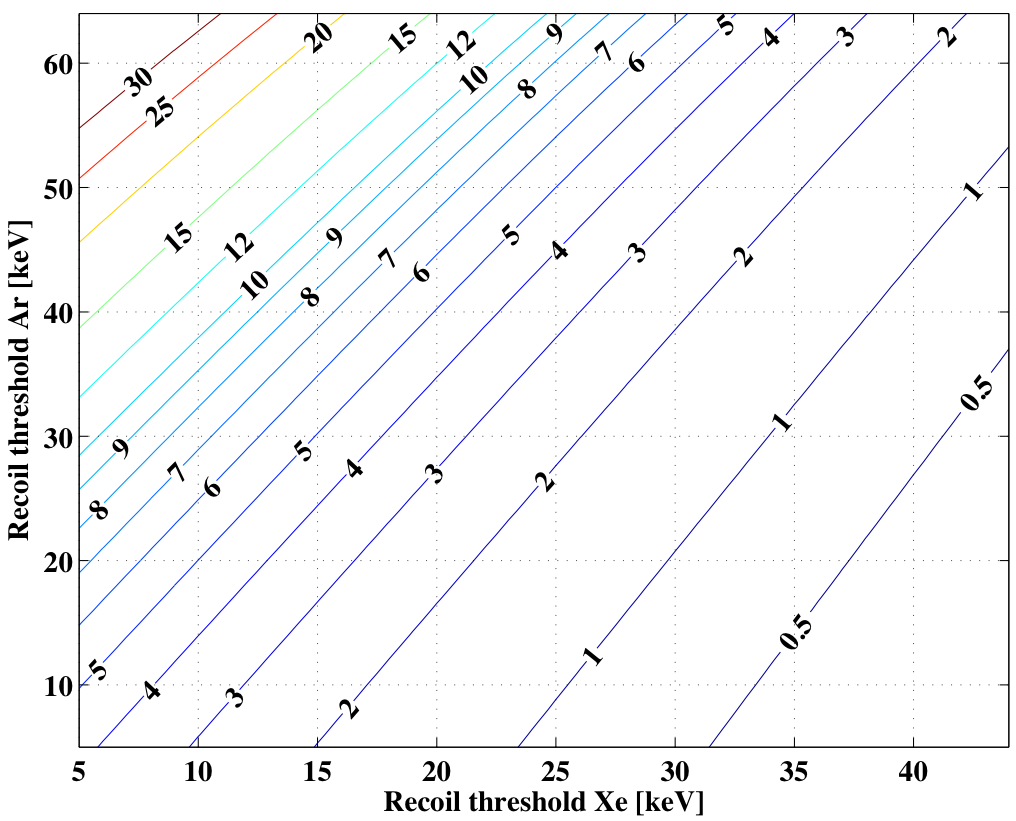}
\caption{\small{(Left): Integrated rate in a LAr (blue) and LXe (black) target as a function of energy threshold for a WIMP-nucleon cross section of 10$^{-45}$cm$^2$ and a WIMP mass of 100\,GeV/c$^2$. (Right): The mass scaling factor between argon and xenon for achieving a similar sensitivity to a standard WIMP that interacts predominantly via coherent scattering on nuclei, as a function of the energy thresholds for nuclear recoils in the respective targets.}}
\label{fig:rates_scaling}
\end{figure}

The final configuration and the size of DARWIN are not yet determined, for these are some of the expected outputs of this study.  There are several considerations that will influence the relative size of the fiducial mass in the two target materials, from detector-dependent quantities such as energy thresholds, to WIMP physics such as type of particle and interaction. Figure  \ref{fig:rates_scaling} shows the integrated rates in a LAr and a LXe target as a function of energy threshold for a WIMP-nucleon cross section of 10$^{-45}$cm$^2$ and a WIMP mass of 100\,GeV/c$^2$. It also shows the mass scaling factor between the two targets for achieving a similar sensitivity to a standard WIMP that interacts predominantly via coherent scattering on nuclei, as a function of the energy thresholds for nuclear recoils in argon and xenon, respectively. 

\begin{figure}[!h]
 \begin{center}
\includegraphics[scale=0.24]{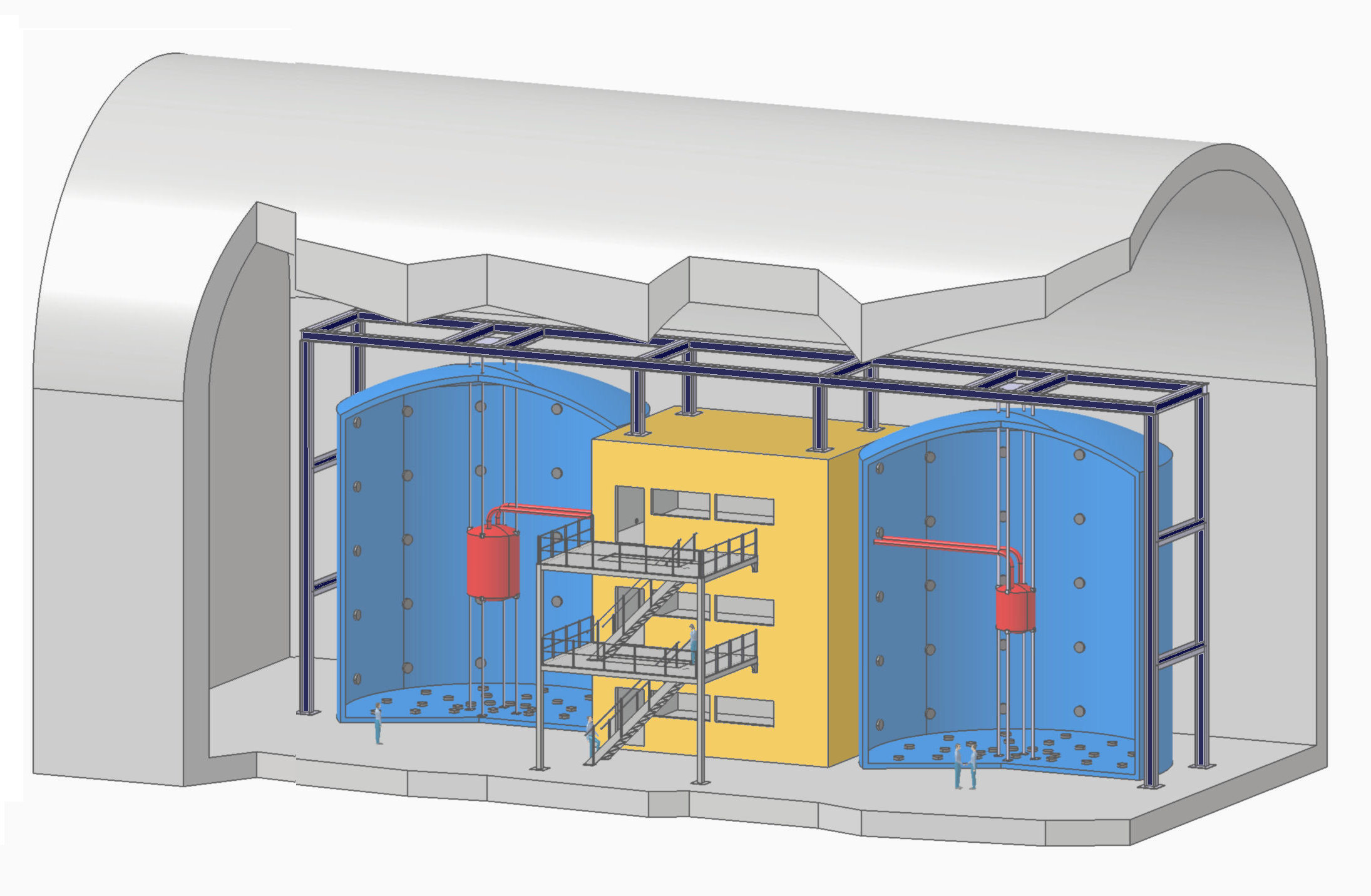}
\caption{\small{A preliminary sketch of the DARWIN facility, which would operate 20\,t (10\,t) and 8\,t (5\,t) of total (fiducial) argon and xenon mass in double-walled cryostats immersed in large water Cerenkov shields.}}
\label{fig:sketch}
\end{center}
\end{figure}

\begin{figure}[!h]
\begin{center}
\includegraphics[scale=0.55]{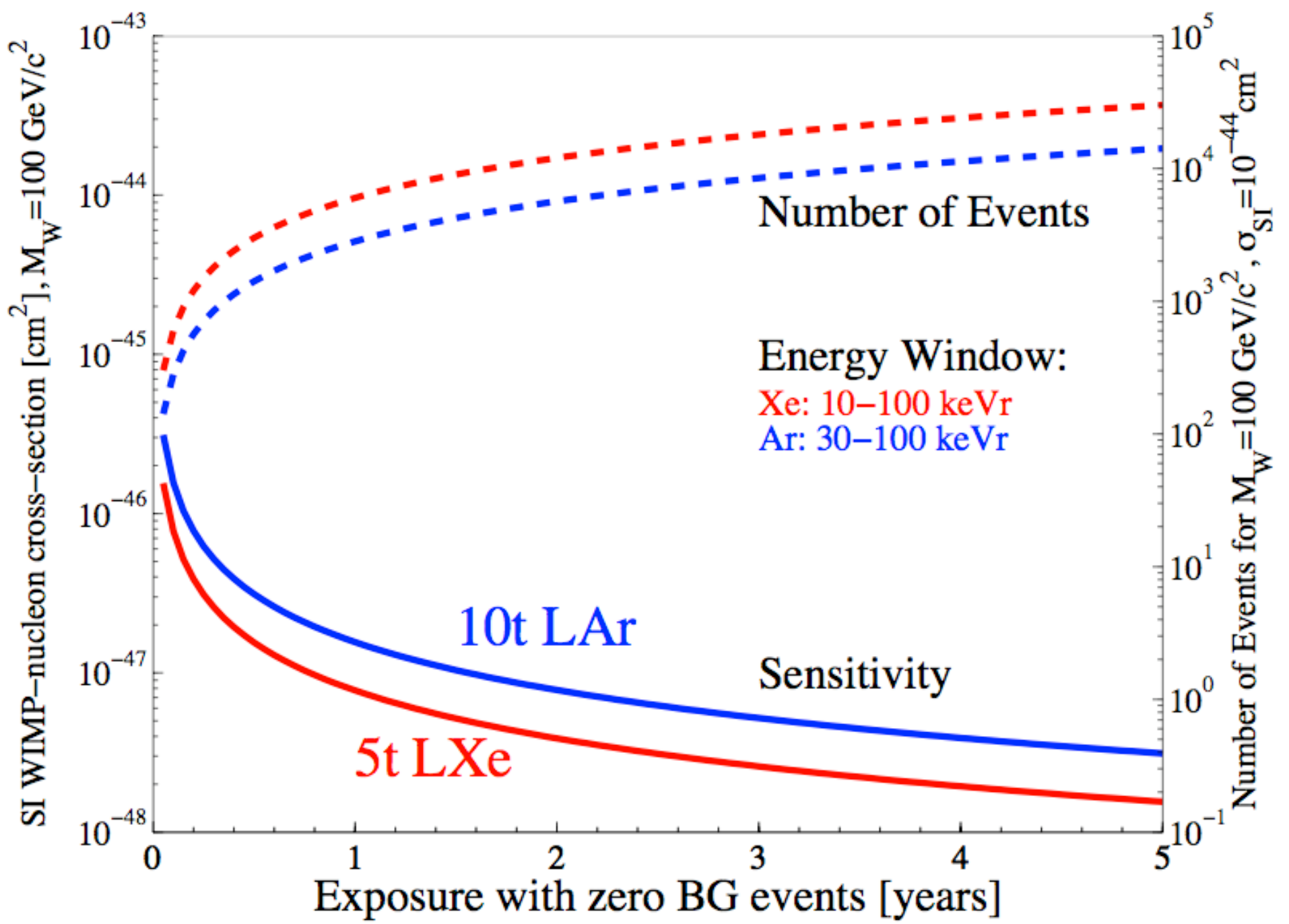}
\caption{\small{The sensitivity to the spin-independent WIMP-nucleon cross section as a function of exposure for 10\,t LAr (blue) and 5\,t LXe (red), for a WIMP mass of 100\,GeV/c$^2$, an energy window of 30-100\,keVr and 10-100\,keVr in LAr and LXe, respectively and zero background events for a given exposure (left y-axis). The dashed lines show the number of events that would be detected for a WIMP-nucleon cross section of 10$^{-44}$cm$^2$ (10$^{-8}$pb) in LAr (blue) and LXe (red) (right y-axis).}}
\label{fig:sensitivity}
\end{center}
\end{figure}

To study the physics reach of the facility, we assume as benchmark scenarios fiducial masses of  10\,t and 5\,t  for the  LAr and LXe components, respectively, corresponding to roughly 20\,t and 8\,t of total argon and xenon mass (a preliminary sketch of DARWIN is shown in Figure \ref{fig:sketch}).  Figure \ref{fig:sensitivity} shows the sensitivity to the spin-independent WIMP-nucleon cross section as a function of exposure for a WIMP mass of 100\,GeV/c$^2$ and an energy window of 30-100\,keVr and 10-100\,keVr in LAr  and LXe, respectively. It also displays the number of events that would be detected for a WIMP-nucleon cross section of 10$^{-44}$cm$^2$ (10$^{-8}$pb) in the same energy windows. The assumptions for LXe are the following:  a raw background of 0.1\,mdru (10$^{-4}$ events kg$^{-1}$day$^{-1}$ keV$^{-1}$), which is a factor of 100 below the current XENON100 background of about 10\,mdru, a 99.9\% rejection of electronic recoils based on the ratio of the charge and light signals, and a 50\% acceptance for nuclear recoils. For LAr, the assumptions are: a raw background of 0.45 dru, with a factor of 10$^8$ rejection of electronic recoils based on pulse shape analysis and the charge-to-light ratio, a reduction of the $^{39}$Ar rate by a factor of 25 relative to atmospheric argon (corresponding to an activity of 40\,mBq/kg for $^{39}$Ar) and a 80\% acceptance for nuclear recoils.

The DARWIN study has officially started in April 2010, and the Technical Design Study is expected to be delivered by early 2013. The letter of intent and the proposal for the construction of the facility  could then be submitted by mid and late 2013, respectively, with the construction and commission phases scheduled for 2014-2015. The period of operation and physics data taking is foreseen for 2016-2020.

In summary, DARWIN is an R\&D and design study for a facility to detect dark matter induced signals by observing the charge and light produced in multi-ton scale liquid noble gas targets, using techniques which have already been successfully proven in 10\,kg-100\,kg prototypes, and which will be studied in ton-scale detectors in the near future.  In conjunction with other WIMP targets, with indirect searches and with the LHC, DARWIN should allow us to learn not only about the WIMP properties, but also about their density and velocity distribution in our local vicinity in the Milky Way. The goal is to probe the spin-independent WIMP-nucleon cross section well below 10$^{-47}$cm$^2$ (10$^{-11}$pb), which is three orders of magnitude beyond the current best limits.

\section{Acknowledgements}
We would like to thank the organizers for a very fruitful and stimulating conference. This work is supported through the first ASPERA common call, from the virtual pot created from contributions from the national funding agencies participating in this call, as well as by the individual institutions participating in the study.

\end{document}